# Entangled messages

Arindam Mitra

Anushakti Abasan, Uttar-Falguni -7, 1/AF, Salt lake,

Kolkata, West Bengal, 700064, India.

**Abstract**: It is sometimes necessary to send copies of the same email to different parties, but it is impossible to ensure that if one party reads the message the other parties will bound to read it. We propose an entanglement based scheme where if one party reads the message the other party will bound to read it simultaneously.

It would be interesting thing if one transmits same or different messages to two parties who will know each other's message simultaneously. One can think that both messages can be transmitted to both parties simultaneously. But there is no guarantee one will show any interest to read other's message and even if they read there is no guarantee they will read simultaneously. Following alternative quantum information processing technique [1-3] we propose an entanglement based scheme where same or different messages can be transmitted to two parties where message cannot be recovered independently.

Four different pairs of entangled sequences of singlet states encode double bits 00,11,01,10. A typical example of the encoding is given below.

$$S^0_i = \{A, B, C, D, E, F, G, H, .........\}$$
$$S^0_j = \{B, F, G, A, E, H, D, C, ..........\}$$

$$S^1_i = \{A, B, C, D, E, F, G, H, .........\}$$
$$S^1_j = \{A, C, G, E, B, D, H, F, ..........\}$$

$$S^0_i = \{A, B, C, D, E, F, G, H, .........\}$$
$$S^1_j = \{F, A, B, D, C, G, E, H, ..........\}$$

$$S^1_i = \{A, B, C, D, E, F, G, H, .........\}$$
$$S^0_j = \{E, C, H, B, E, A, C, D, ..........\}$$

The same pair of letters denotes an EPR pair. The singlet state may be written as

$$|\Psi_-\rangle_{i,j} = \frac{1}{\sqrt{2}}\left(|+\rangle_i|-\rangle_j - |-\rangle_i|+\rangle_j\right)$$

where i,j denotes the position of an EPR pair in $S_i$ and $S_j$ and "+" and "−" denote two opposite spin components. There are three parties Alice, Bob and Sonai. Three parties share the sequence codes. Bit value associated with $S_i$ is transmitted to Bob and bit value associated with $S_j$ is transmitted to Sonai. Alice sends $S_i$ and $S_j$ to Bob and sonai respectively. Bit value can be recovered in the following way.

1. Bob and sonai measure spin of their particles in z-direction.
2. Bob and sonai disclose their results to each other.
3. Both arrange the two sets of results according to a probable sequence code, say $S^0_i$ and $S^0_j$. If they get 100% EPR correlated data then both get "0". If 100% EPR correlation is not found, they again arrange the results according to another probable sequence code, say $S^1_i$ and $S^1_j$. If 100% EPR correlation is found then both get "1". If 100% EPR correlation is not found they again arrange according to another probable sequence code, say $S^0_i$ and $S^1_j$. If 100% EPR correlation is found then Bob gets "0" sonai "1". If 100% EPR correlation is not found they again arrange according to the remaining probable sequence code $S^1_i$ and $S^0_j$. This time 100% EPR correlation will be certainly found and Bob will get "1" and sonai "0".

Both will know each other's message almost simultaneously. But one can delay in revealing the results. This cheating strategy can be foiled if none of them reveals all the results at a time but reveals results alternatively. That is, if Alice reveals a result Bob will reveal the next result. In this way, both can recover bit value simultaneously with same statistical confidence level.